\documentclass[pre,aps]{revtex4}
\usepackage{graphicx} 
\begin{document}
\title{Stability of supercooled binary liquid mixtures}
\author{S{\o}ren Toxvaerd, Ulf R. Pedersen, Thomas B. Schr{\o}der, and Jeppe C. Dyre}
\affiliation{DNRF centre  ``Glass and Time,'' IMFUFA, Department of Sciences, Roskilde University, Postbox 260, DK-4000 Roskilde, Denmark}
\date{\today}

\begin{abstract}
Recently the supercooled Wahnstrom binary Lennard-Jones mixture was partially crystallized into ${ \rm MgZn_2}$ phase crystals in lengthy Molecular Dynamics simulations. We present  Molecular Dynamics simulations of a modified Kob-Andersen binary Lennard-Jones mixture that also crystallizes in lengthy simulations, here however by forming pure fcc crystals of the majority component. The two findings motivate this paper that gives a general thermodynamic and kinetic treatment of the stability of supercooled binary mixtures, emphasizing the importance of negative mixing enthalpy whenever present. The theory is used to estimate the crystallization time
in a Kob-Andersen mixture from the crystallization time in a series of relared  systems.  At T=0.40 we estimate this time to be 5$\times 10^{7}$ time units ( $\approx 1. ms$). A new binary Lennard-Jones mixture is proposed that is not prone to crystallization and faster to simulate than the two standard binary Lennard-Jones mixtures; this is obtained by removing the like-particle attractions by switching to Weeks-Chandler-Andersen type potentials, while maintaining the unlike-particle attraction.
\end{abstract}

\vspace*{0.7cm}

\maketitle

\newcommand{\tf} {T_{\textrm{fus,A}}}
\newcommand{\hf} {H_{\textrm{fus,A}}}
\newcommand{\sfu} {S_{\textrm{fus,A}}}
\newcommand{\hm} {H_{\textrm{mix,A}}}
\newcommand{\sm} {S_{\textrm{mix,A}}}
\newcommand{\dgm} {\Delta G_{\textrm{mix,A}}}
\newcommand{\dhm} {\Delta H_{\textrm{mix,A}}}
\newcommand{\dsm} {\Delta S_{\textrm{ideal mix,A}}}
\newcommand{\dsmix} {\Delta S_{\textrm{ideal mix}}}
\newcommand{\dsmid} {\Delta S_{\textrm{ideal mix,A}}}
\newcommand{\dgf} {\Delta G_{\textrm{fus,A}}}
\newcommand{\dhf} {\Delta H_{\textrm{fus,A}}}
\newcommand{\dsf} {\Delta S_{\textrm{fus,A}}}
\newcommand{\dcf} {\Delta C_{p,\textrm{fus,A}}}
\newcommand{\dta} {\Delta T_{\textrm{fus,A}}}
\newcommand{\ma} {\mu_{\textrm{A}}}
\newcommand{\xa} {x_{\textrm{A}}}
\newcommand{\xb} {x_{\textrm{B}}}
\newcommand{\dt}{\Delta T}

\section{Introduction}

As computers get faster, simulations of the highly viscous liquid phase preceding glass formation become increasingly realistic. In this context it is nice to have a  standard model system, something like the Ising model for critical phenomena. For several years  binary Lennard-Jones (BLJ) mixtures
have served this purpose -- in particular the Wahnstr{\"o}m (Wa) and Kob-Andersen (KA) systems
\cite{wahn,kob}, because they are easy to simulate and were never found to crystallize.
The KA BLJ consists of two types of Lennard-Jones particles, 80\% large (A)
particles and 20\% small (B) particles. The KA  potentials are modifications of the potentials
devised by Weber and Stillinger \cite{web}, who constructed the pair potentials for the binary
mixture based of physical-chemical data for the Ni$_{80}$P$_{20}$ alloy. The Kob-Andersen potentials
describe a strongly non-ideal mixture due to an AB attraction that is three times stronger than the
BB attraction. This ensures a large negative mixing enthalpy (and energy), which  as detailed below
suppresses crystallization into pure A crystals.

Recently, the Wa system was shown to crystallize in lengthy computer runs \cite{arxiv}. 
This motivated the present paper that has three purposes. First, we  review the general theory
of thermodynamic and kinetic stability of supercooled binary mixtures and derive a relation for
nucleation times of the solvent in binary mixtures (Sec. II). Secondly, we detail
the crystallization of  modified Kob-Andersen (MKA) type systems. Their nucleation times are
used to estimate the nucleation time of the KA system to be of the order 
 roughly $0.1$ milliseconds in Argon units (Sec. III); the crystallization of the
Wa liquid  will be described elsewhere \cite{urp}. Finally, we suggest a new KA-type  BLJ
system that is faster to simulate and which is even less prone to   crystallization
than the KA system. The idea is to keep the KA system's large negative mixing
enthalpy, but remove the AA and BB attractions by adopting Weeks-Chandler-Andersen (WCA) type potentials
between like particles (Sec. IV). Section V gives a brief discussion.

\section{General treatment}

\subsection{ Freezing-point depression}

Consider a binary mixture of $N_A$ solvent (A) particles and $N_B$ solute (B) particles, with total number of particles $N=N_A+N_B$. The thermodynamic stability of this system against crystallization is
expressed by the melting temperature of a pure A crystal, $\tf$, as a function of the concentration of A particles.
The latter quantity is  conveniently expressed in terms of the fraction $x_{\textrm{B}}=N_B/N$ 
 ($\xa+\xb=1$). We consider the usual case of externally controlled temperature $T$ and pressure $p$.
At the melting temperature of the pure A crystalline phase in the AB mixture, the chemical potential of
the A particles in the crystal equals the chemical potential of the A particles of the liquid mixture.
The chemical potential is the Gibbs free energy per particle. The change in  Gibbs free
energy per A particle at melting, $\Delta G_{\textrm{trans,A}}$ can be divided into two terms,
$\dgf$ and  $\dgm$, where  $\dgf$ is the change in Gibbs free energy per A particle upon melting
an A crystal into pure A liquid, and $\dgm$ is the change in Gibbs free energy per A particle
going from pure to mixed liquid. The melting temperature $\tf$ is determined by

\begin{equation}\label{fuseq}
\Delta G_{\textrm{trans,A}} = \dgf(\tf)\ + \dgm(\tf)\,=\,
0\,.
\end{equation}
The case of pure A ($\xb=0$) will be denoted by an asterisk, thus $\tf^*$ is the melting point of the pure A crystal into pure A liquid. In this case $\Delta G_{\textrm{trans,A}} = \dgf^*$ and since $G=H-TS$, one has

\begin{equation}
\dhf^*-\tf^*\dsf^*=0,
\end{equation}
where here and henceforth all thermodynamic quantities are per A atom.
When deriving the standard expression for the freezing point depression 
one makes the approximation that crystal and liquid have the same specific heats,
i.e., that $\dhf$ and $\dsf$ are temperature independent, or from Eq. (2)

\begin{equation}
\dhf-\tf^*\dsf=0.
\end{equation}
In the general case where $\xb\neq 0$ one has from Eq. (1) 

\begin{equation}
\dhf-\tf\dsf=-\dgm,
\end{equation}
and by using Eq. (3) to  eliminate $\dsf$ in Eq. (4) one obtains

\begin{equation}\label{dteq}
\frac{\dta}{\tf^*}\,=\,
-\frac{\dgm}{\dhf}\,
\end{equation}
for the melting-point depression $\dta=\tf^*-\tf$.
In the standard textbook treatment one assumes that $\dhm=0$ and that the mixing entropy is ideal,
i.e., that the mixing entropy divided by $N=N_A+N_B$ is given by
$\dsmix=-k_B[\xa\ln(\xa)+\xb\ln(\xb)]$.
This expression separates into a contribution from the A particles and one from the B particles.
Per A particle we thus have $\dsm=-k_B\ln(\xa)=-k_B\ln(1-\xb)$.
Under these assumptions Eq. (5)
for $\xb \rightarrow 0$ reduces to the well-known expression

\begin{equation}\label{textbook}
\frac{\dta}{\tf^*}\,\cong\,
\frac{k_B\tf}{\dhf^*}\,\xb\,.
\end{equation}
More generally, $\dhm=0$ does not apply -- in fact for the Kob-Andersen liquid the mixing
enthalpy is large (and negative) and this term cannot be ignored.
For small concentrations of the solute, $\xb$, the mixing entropy is still
given by $\dsmid$. For this more general case Eq. (\ref{dteq}) becomes

\begin{equation}\label{dteqnew}
\frac{\dta}{\tf^*}\,=\,
-\frac{\dhm+k_B\tf\ln(\xa)}{\dhf}\,.
\end{equation}
A negative mixing enthalpy (i.e. exotherm)  clearly implies a further melting-point depression
beyond that of the traditional treatments. If there is a large negative mixing enthalpy, this effect cannot be ignored.

\subsection{ Crystallization}

The observed stability of a supercooled liquid mixture depends not only on its absolute (thermodynamic)
stability against crystallization, but also on kinetic effects. The less supercooled the liquid is,
the larger is the critical nucleus. This means that it takes longer time before a critical nucleus is generated by a thermal fluctuation, i.e., the supercooled liquid is more stable. At a given temperature the supercooled liquid is more stable the more negative $\dgm$ becomes. To see this divide the creation of a crystal nucleus of pure A particles in the mixture

\begin{equation}
N_{\textrm{A}}({\rm mix}) \rightarrow N_{\textrm{A}}({\rm crystal},x_{\textrm{A}}=1)\\
\end{equation}
into two steps:

\begin{eqnarray}
N_{\textrm{A}}({\rm mix})& \rightarrow & N_{\textrm{A}}({\rm liquid},x_{\textrm{A}}=1)\nonumber\\
N_{\textrm{A}}({\rm liquid},x_{\textrm{A}}=1)& \rightarrow  & N_{\textrm{A}}({\rm crystal},x_{\textrm{A}}=1)\,.
\end{eqnarray}
The first reaction is a composition fluctuation from the mixture to a pure liquid domain of A particles, the second is crystallization. The crystallization is (qualitatively) described by classical nucleation theory (CNT). In its simplest formulation the number of particles, $N^*$, in the critical nucleus is given \cite{Laak} by

\begin{equation}
N^*=\frac{32 \pi \gamma_{\infty}^3}{3\rho_{c}^2 \Delta \mu^3} \,,
\end{equation}
where $\gamma_{\infty}$ is the solid-liquid surface tension, $\rho_{c}$ is the crystal number density
and $\Delta \mu \equiv \Delta \ma$  is the  change in Gibbs free energy per A particle by going from
 crystal to  liquid. When the creation of a pure A crystal takes place in a mixture, the lowering of Gibbs free energy by the crystallization process is reduced by $\delta \mu=-\dgm=-\dhm+\tf\dsm \approx -\Delta_{{\textrm{mix,A}}}u_{pot}- k_B\tf\ln(\xa) $. Since the size of the critical nucleus $N^*$ varies as $N^*\propto (\Delta\ma)^{-3}$,  when $\Delta\ma$ is replaced by $\Delta\ma-\delta \mu_{\textrm{A}}$ to lowest order the change in critical nucleus size $\delta N^*$ is given by $\delta N^*/N^*=3\delta\ma/\Delta\ma$, thus

\begin{equation}\label{11}
\frac{\delta N^*}{N^*}\,=\,
-\,3\,\frac{\Delta_{{\textrm{mix,A}}}u_{pot}(\xa)+k_B\tf\ln(\xa)}{\Delta \ma}\,.
\end{equation}
The  stability criteria indicate that if a binary mixture crystallizes into, e.g., an AB-type crystal,
a negative mixing energy will enhance the tendency of crystallization by
decreasing the size of the critical nucleus. Confirming this, we find in our simulations that
increasing the fraction of B particles to $x_{\textrm{B}}$=0.5 for KA-type
BLJ mixtures results in systems that  quickly crystallize into an AB
(CsCl structure) crystal. This is consistent with the results of Fernandez
and Harrowell \cite{far03}, who found that the $T=0$ equilibrium phase of
the KABLJ mixture consists of coexisting pure A (FCC) and AB (CsCl structure) crystals.

Classical nucleation theory  can be used to estimate the crystallization time, $\tau$, 
of the solvent in a supercooled  binary mixture at  temperature $T < T_{\textrm{fus,A}}$, given
that the crystallization time  is known for  "similar" systems
that are more prone to crystallization. In the derivation below
we consider crystallization in systems with different strength of solvation energy, given by the attraction
between solvent and solute particles (Sec. III).\\ 
According to CNT the  nucleation rate in CNT is given by  \cite{Ox},\cite{wolde}

\begin{equation}
k_{\textrm{CNT}}=Zf_e(N^*)\rho_ce^{-\beta \Delta G(N^*)},
\end{equation}
where $Z$ is the  "Zeldovich" factor,
$f_e(N^*)$ is the "forward rate" per particle  of the growth of the critical nucleus with  $N^*$ particles, and
$\Delta G(N^*)$ is the Gibbs free-energy barrier for the critical nucleus,

\begin{equation}
  \Delta G(N^*)=\frac{16 \pi \gamma^3 }{3 \rho_c \Delta \mu^2}.
\end{equation}
We now consider the ratio between  the nucleation
rates $k_j/k_i$ of two similar systems.
This ratio consists of a pre-exponential factor and the difference 
between the two free-energy barriers in an exponential. The later difference

\begin{equation}
\Delta G_i(N^*)-\Delta G_j(N^*)=\Delta G_i(N^*)\left(1-(\frac{\Delta \mu_i}{\Delta \mu_j})^2\right)
\end{equation}
(assuming that the surface tensions of the critical nuclei are the same.)
In the case of crystallization of the solvent from a binary mixture with particle fraction $x_{\textrm{A}}$
of the solvent
particles, the change in free energy per A-particle is given by

\begin{equation}
\Delta \mu =-\Delta \mu_{fus}+\Delta_{\textrm{mix,A}}u_{pot} - T \Delta S_{\textrm{mix,A}}.
\end{equation}
It is not necessary to assume ideal mixing;  only that the entropy per A-particle is the same
for the binary mixtures with different strength of solvation forces.
The difference in  Gibbs-free energy change per particle for crystallization in the two
systems only differs by the potential energy per A particles in the two mixtures (with the same solute concentration)

\begin{equation}
\Delta \mu_j =
\Delta \mu_i -\Delta u_{ij}(x_{\textrm{A}}),
\end{equation}
where  $\Delta u_{ij}(x_{\textrm{A}}) \equiv u_{\textrm{A,pot,i}}(x_{\textrm{A}}) - u_{\textrm{A,pot,j}}(x_{\textrm{A}})$.
Using the relation between $\Delta \mu_j$ and  $\Delta \mu_i$  one obtains

\begin{equation}
 \Delta G_i(N^*)-\Delta G_j(N^*)  \approx -\Delta G_i(N^*)\frac{2 u_{ij}}{\Delta \mu_i}.
\end{equation}

The (forward) rate at which a crystal grow, $f_e(N^*)$  is proportional to the area of the 
crystal and to the mean speed of the particles that hits the crystal surface.
The speed is proportional to the self-diffusion constant, $D$,
of solvent particles in the mixture.\\
The  self-diffusion constant for a supercooled highly viscous liquid
decreases faster than an Arrhenius expression with the degree of supercooling,
and with  increasing solvation energy.  In contrast, 
the pre-exponential factor $Z$ and the area of the critical nucleus vary only algebraic with
$\Delta \mu(T)$, so to first order 

\begin{equation}
\frac{Z_j f_{e,j}(N_j^*)}{Z_i f_{e,i}(N_i^*)} \approx \frac{D_j(x_{\textrm{A}})}{D_i(x_{\textrm{A}})}, 
\end{equation}
and the ratio between the nucleation rates, $k_j/k_i$ or mean nucleation time $\tau_i/\tau_j$
is 

\begin{equation}
\frac{k_j}{k_i}=\frac{\tau_i}{\tau_j} \approx \frac{D_j}{D_i}e^{- \beta \Delta G_i(N^*)\frac{2 u_{ij}}{\Delta \mu_i}}
\end{equation}
from which the nucleation time for a  system can be estimated from  similar systems
if  the nucleation times, $D$ and $ u_{\textrm{A,pot}}$ are known.\\
We  studied  four modified Kob-Andersen (MKA) systems (Sec. III) to estimate the mean nucleation time of the 
KA system of $N_{\textrm{A}} =800$ and $N_{\textrm{B}}=200$ particle at the particle density
$\rho=1.2$ and temperature $T=0.40$.

\section{Crystallization in binary mixtures }

\subsection{ Wahnstr{\"o}m,  Kob-Andersen, and  modified Kob-Andersen binary Lennard-Jones mixtures}

The simplest case of a realistic mixture is that given by the semi-empirical Lorentz-Berthelot (LB) rules
\cite{Row} relating the energy parameters $\sigma$ and $\epsilon$ in the pair potentials between the
different species of the AB mixture as follows:

\begin{eqnarray}
\sigma_{\textrm{AB}}=(\sigma_{\textrm{AA}}+\sigma_{\textrm{BB}})/2\\
\epsilon_{\textrm{AB}}=\sqrt{\epsilon_{\textrm{AA}}\epsilon_{\textrm{BB}}}\,.
\end{eqnarray}
These rules apply for atoms with  weak dispersion attractions \cite{Hirs}, and they work well for simple mixtures of for instance noble-gas atoms \cite{Row}, i.e., for interactions between spherically symmetric atoms with unperturbed valence-electron orbitals.
There are two standard models for highly viscous supercooled binary mixtures,
the model introduced by Wahnstr{\"o}m (Wa) \cite{wahn} and the model by
Kob and Andersen (KA) \cite{kob}. Both models involve Lennard-Jones potentials (LJ).

\begin{equation}
u_{\textrm{LJ}}(r_{ij})=4\epsilon_{ij}\left[\left(\frac{\sigma_{ij}}{r_{ij}}\right)^{12}-
\left(\frac{\sigma_{ij}}{r_{ij}}\right)^6\right]
\end{equation}
between particle $i$ and  $j$. The Wa model obeys the LB-rules by having 

\begin{equation}
\sigma_{\textrm{AB}}=(\sigma_{\textrm{AA}}+\sigma_{\textrm{BB}})/2,\\
\end{equation}
with $\sigma_{\textrm{BB}}=1/1.2\sigma_{\textrm{AA}}$ and $\epsilon_{\textrm{AB}}=\epsilon_{\textrm{AA}}=\epsilon_{\textrm{BB}}=1$. 
The KA model strongly disobeys the LB-rules: The LJ-potential parameters for the KA mixture are 
$\sigma_{\textrm{AB}}=0.8 \sigma_{\textrm{AA}}$, $\sigma_{\textrm{BB}}=0.88 \sigma_{\textrm{AA}}$, 
$\epsilon_{\textrm{AB}}=1.5 \epsilon_{\textrm{AA}}$, and $\epsilon_{\textrm{BB}}=0.5 \epsilon_{\textrm{AA}}$.
The fact that the binding energy of the small
solute particle to a solvent particle is three times larger
than the binding energy between two solute particles results in a tendency for the small
and mobile solute B particles to ``glue'' to the A particles.
The KA mixture has a significant non-ideal mixing energy and it is very stable against crystallization,
much more so  the Wa mixture. So far the KA system has not been crystallized.

We  estimated  \cite{cal} the melting-point temperature at constant pressure
$(p \sigma_{{\textrm{AA}}}^3/\epsilon_{{\textrm{AA}}}$=10) for small solute
concentrations, $\xb$,   using Eq. (7)  for the different models. The result
is shown in Fig. 1. The melting-point depression by mixing is traditionally
given at constant pressure. Most MD simulations are, however, performed for
isochores where $T_{{\textrm{fus,A}}}(x_{{\textrm{B}}})$ depends on the partial volumes of the species in the mixture.

As can be seen from the figure the negative mixing energy lowers the melting point
temperature significantly. The Wa model is indeed more prone  to crystallization
than the other models and recently Pedersen and co-workers reported crystallization
of the Wa model after lengthy computer runs \cite{urp}. The crystal is the
${\rm MgZn_2}$ phase consisting of particles in the ratio 1:2, different from
the 1:1 ratio defining the Wahnstr{\"o}m BLJ. Thus concentration fluctuations
precede (or at least correlate with) crystal formation.

The KA model has served as a standard computer  model for investigations of
long time behavior   of highly viscous liquids. With the continuous growth of
computer capability it is now  realistic to simulate viscous system over times which
approach $ms$. It is therefore useful to know
whether  this system will crystallize within this time interval. According to the previous
section one can estimate the nucleation time, $\tau(\textrm{KA})$, in the KA system  from the the
nucleation times $\tau(\textrm{MKA})$ in "similar" systems. The similar systems must be such that the 
distribution of the solvent particles are the same, but with a weaker mixing energy, given
by the mean potential energy of the solvent particles in the mixtures.
We  performed four sets of simulations with a scaled energy of attraction
of the solute particle, B, given by the energy parameter:

 \begin{eqnarray}
 \epsilon_{\textrm{AB}}(\textrm{MKA})=\epsilon_{\textrm{AB}}(\sigma_{\textrm{AB}}/
  \sigma_{\textrm{AA}})^n\\  
   \epsilon_{\textrm{BB}}(\textrm{MKA})=\epsilon_{\textrm{BB}}(\sigma_{\textrm{BB}}/\sigma_{\textrm{AA}})^n, 
\end{eqnarray}   
with $n$=1.5, 2, 2.5 and 3. In the following the four systems are labled MKA1.5, MKA2, MKA2.5 and
MKA3, respectively.
The scaled energies of the solute and its interaction  with the solvent particles do not affect
 the distribution of the solvent particles. 
 Fig. 2 shows the radial distribution functions, $g_{\textrm{A,A}}(r)$
 and $g_{\textrm{A,B}}(r)$ for the  distribution of A-A interactions and A-B interactions.
 Whereas $g_{\textrm{A,A}}(r)$  is only  little affected by the  weaker binding to the solvent particles, 
 $g_{\textrm{A,B}}(r)$ shows a reduced probability for a direct attachment of the B-particles. The
  corresponding mean potential energy of the solvent particles
  in the mixture at   $\rho =1.2$ and $T=0.40$ are shown in Table I.  The reduced energy of the solute
 particles leads to a corresponding reduction of the (negative) potential energy, $u_{\textrm{A,pot}}(x)$.

\subsection{Crystallization of a modified Kob-Andersen binary mixture}
In order to investigate the crystallization properties of the KA and MKA mixtures
we performed molecular dynamics
simulations of 1000 (occasionally 10000) particles of the  MKA-version  
 in both the NVE and  the NVT ensembles. The idea is that the KA system
 crystallizes
by a mechanism similar to that of the modified system.

The standard time-reversible leap-frog (NVT) algorithm \cite{tox1} was used. In the following
 data are given in unit length  $\sigma_{\textrm{AA}} $, unit energy $\epsilon_{\textrm{AB}}$, and
unit time $\sigma_{\textrm{AA}} \sqrt{m/\epsilon_{\textrm{AA}}}  (\approx 2. \times 10^{-12} s$ in Argon units).
 The software used has been described elsewhere \cite{toxsoft};
it utilizes a double sorting of neighbor particles that makes
it possible to simulate 1 $\mu$s within 2-3 days of computing on a standard computer.

With the MKA2 system we did not  detected  crystallization for temperatures above 0.45.
In the temperature interval $[0.39,0.45]$
a  drop in pressure taking place typically after $\approx 10^6$ time units indicates that 
 the system phase separates   such that the A particles form fairly large regions
with no B particles present. Linked to this phase separation is a crystallization of the A particles.
Ten simulations with the MKA2 system were performed in the $[0.39,0.45]$ temperature interval; after $2 \times 10^6$ time
steps ($\approx 4 \mu$s) eight of these ten simulations phase separated with crystallization of the A particles.
Fig. 3 shows a representative example of the crystallization, giving the positions of the particles after
crystallization at T=0.40 (NVT-MD). There is a large region of pure A particles showing clear crystalline order.

\subsection{ Estimation of the crystallization time for the Kob-Andersen binary mixture}

 According to the theory in Section II B, the nucleation time in a
 KA mixture can be estimated from the nucleation time in similar systems.
We  performed  MD simulations for  four systems with different scaled solute-solute and
solute-solvent energy, given by the power, $n$=1.5, 2, 2.5 and 3, respectively (see Eq. (24)-(25)). The mean 
nucleation times, $\tau$, obtained  from a maximum likelihood estimate \cite{xxx}
based on  twelve independent
simulations for each of  the four MKA mixtures at  $T=0.40$, $\rho=1.2$ and $x_{\textrm{A}}=0.8$ are shown in 
Table I together with the corresponding  values of  the self diffusion constants, $D_\textrm{A} $, and
the mean potential energies, $u_{\textrm{A,pot}}$, per A-particles in the binary mixtures.

In principle the estimation of nucleation times can be performed in two different
ways: 1) By  linear extrapolation of $\ln \tau(n)$ as a function of  $n$;  2) 
 From  $\ln[ \tau_i D_i(u_{i})]$, according to Eq. (19), as a function of the 
 mean potential energy of the nucleating solvent particles in the binary mixture.
The data in Table I is used for the  estimations of the crystallization time for the KA system.
 Method 1 predicts a mean nucleation time for the classical KA mixture  of the order
 5$\times 10^{7}$ time units ( $.1 ms$ Argon units).  Concerning method 2 we notice that,
 although the uncertainty in the nucleation time $ \tau_i$ is large,
 the variation of  $\ln [ \tau_i D_i(u_{i})]$
is relative  small  and in the interval $[2.5,3.5]$ for the four MKA systems. This indicates
that the main effect of the resistance against crystallization in the systems arises indirectly from
the increased viscosity by increasing the
solvation energy,  not directly from the solvation energy itself, given by the
the term $u_{i}$ in Eq. (19). Using method 2 and with $\ln [ \tau_i D_i(u_{i})] \approx 3.83$ gives a nucleation
time  5$\times 10^{7}$.\\ 

Our conclusion, based on the two methods,  is that the KA mixture will nucleate  roughly 
within  5$\times 10^{7}$ time units, a time interval which
is  realistic for computer simulations in the near future.\\

\section{A new binary model system }

With the general theory for melting point depression and nucleation in mind it is possible to change 
existing  binary models in such a way that they are faster to simulate and even less  prone to crystallization.
The idea is to construct a new mixture with only repulsive forces between A-A and B-B particles,
but where the attraction between A- and B-particles that is crucial for preventing crystallization
 is maintained. With Lennard-Jones (LJ)
potentials between particle  $i$ and $j$ the modified potentials are

\begin{eqnarray}
&u_{ij}(r_{ij})= \left \{ \begin{array}{ll}
u_{\textrm{LJ}}(r_{ij})-u_{\textrm{LJ}}(r_{ij}(\textrm{cut})), &
r_{ij}< r_{ij}(cut) \\
0, & r \geq  r_{ij}(\textrm{cut}), \\
\end{array} \right.
\end{eqnarray}
with

\begin{eqnarray}
r_{i,i}(\textrm{cut})=2^{1/6}\sigma_{ii}\\
r_{\textrm{AB}}(\textrm{cut})=2.5 \sigma_{\textrm{AB}}
\end{eqnarray}
 {\textit{A priory}} it is to be expected \cite{wca}
that the structure of the mixture only deviates marginally from the corresponding  KA or MKA models
detailed above, because the repulsive parts of the potential interactions
are not changed. The new binary model is more
stable,
and we have so far not been able to crystallize it (for a particle fraction $x_{\textrm{B}}=0.2$).
The new system(s) is  three times faster to simulate than the corresponding KA or  MKA mixtures.

The modification can be applied to any (MKA and KA) of the BLJ mixtures investigated above. We 
used it for the MKA2 system with $n=2$.
The new system was cooled and  equilibrated at a new and lower temperature  until
the diffusion constant remained constant. The system was then simulated
5$\times 10^{7}$ time units (corresponding to $\approx 0.1$ $ms$) for  nine temperatures
in the interval $T \in [0.25,0.40]$. There are two reasons for the resistance
against crystallization by supercooling. First the melting point temperature,
$T_{\textrm{fus,A}}^*$, of the pure solvent particles is lowered significantly
(see Fig. 1).  Secondly, the relative importance of the mixing energy, $\Delta_{{\textrm{A,mix}}}u_{pot}(x_{{\textrm{B}}})$ 
is bigger due to a stronger violation of the LB-rules for the energy and thereby a lower melting-point
temperature and an increase of the size of the critical fcc-nucleus.

The question arises, however, whether the new system is supercooled or simply a highly viscous
equilibrium liquid. As mentioned most MD simulations are performed at constant density.
Thus the result shown in Fig. 1 for $p=10$ cannot  be used to obtain the melting-point
temperature  in the new system at
$ \rho$=1.2 and $x_{\textrm{B}}$ =0.2, and to determine whether
the system is in  a supercooled  state.
The MKA2 mixture crystallizes into
a pure A-particle fcc-structure for small particle fraction,
$x_{\textrm{B}}$, and into a CsCl (A,B) crystal for  $x_{\textrm{B}} \approx 0.5$.
In between there might be other stable crystal arrangements \cite{wales1}.
The relative strength of the binding energy between A- and B particles
 increases when the attraction between A,A- and B,B particles is removed, but
with unchanged attraction between A- and B particles. This change stabilizes
the CsCl-like crystal relative to the liquid, and the melting point temperature at 
$x_{\textrm{B}}$ =0.5  increases, whereas the melting-point temperature of the fcc crystal
for small  $x_{\textrm{B}}$ is decreased (as shown in Fig. 1). 
The qualitative change in melting point temperature  by the modification at constant pressure
is shown in Fig. 4.

There is
a simple computer experiment to investigate whether the new system at  $ \rho$=1.2
and $x_{\textrm{B}}$ =0.2 is in a supercooled state at the low temperature:
First we  crystallized the new system at  particle fraction $x_{\textrm{B}}$=0.5.
Then the system with  $ \rho$=1.2 and $x_{\textrm{B}}$ =0.2 was grafted with a
CsCl crystal of 200 A and  200 B-particles, taken from the  $x_{\textrm{B}}$=0.5
system and surrounded by 600 A particles at a density $\rho =1.2$. The
melting-point temperature of this crystal was directly determined by heating/cooling.
The melting point temperature of the CsCl grafted system at  $x_{\textrm{B}}$ =0.2 and $ \rho$=1.2 is
$T_m \approx 0.32$, and the new system is indeed a supercooled
highly viscous liquid at the low temperatures investigated. For  $T=$0.30 and below the system
with the AB crystal of 400 particles remained; but
the remaining A-particle crystallized into a fcc-structure. The configuration is shown in Fig. 5.
The computer experiment demonstrates that the supercooled system is below what corresponds
to the eutectic  temperature in a constant pressure system. The actual
phase diagram, however, might be even more complicated than given here \cite{wales1}.

A further advantage of the new system is that it is much faster
to simulate by using a two-step sorting of nearest neighbor \cite{toxsoft}. In addition to that one can  use
a larger value of the time increment in the MD simulation due to
the lower temperature, so not only does the mixture not crystallize;
but it can also be cooled further down under equilibrium condition
(see Figs. 7-9) and followed for  longer times.

The structure of the new mixture is, as expected, almost the same as in the
KA and the modified KA mixtures.
Fig. 6 show the radial distribution functions, $g_{\alpha,\beta}(r)$ for the new mixture and for the
corresponding MKA2 mixture.
The similarity is well known: The so-called WCA- system \cite{wca} with  only repulsive
LJ forces has been used in perturbation theories for dense liquid.
As can be seen from the figure only  the distribution, $g_{\textrm{B,B}}(r)$ of the B-particles is affected.

The MKA2 and the new mixture were cooled down from T=1.5. The MKA mixtures crystallize
as mentioned in the temperature interval  T $\in [0.39,0.45]$,
but it is possible to quench the MKA2 down to T=0.375
and still determine the self diffusion constant, D. The new mixture was
cooled down to T=0.25. At each (low) temperature the system was followed
5$\times 10^{7}$ time units $(\approx$ one 0.1 $ms$).
The   diffusion constant, at T=0.25, for the A particles is $D_{\textrm {A}} \approx 1.0\times 10^{-8}$.
The B particles behave in a similar way, but as the temperature is decreased, the ratio $
D_{\textrm {B}}/D_{\textrm {A}}$ changes from 1.7 for T=1.5 to 9.7 for T=0.25.
The values  of the self-diffusion constants $D(T)$
were obtained from the slopes of the mean-square displacements as  function of  time.
At  low temperatures the ballistic and diffusive time regimes are very well separated;
a  log-log plot of  the mean-square displacement 
for  different temperatures  shows this separation (Fig. 7).

The diffusion constants for A  and B particles  are shown in Fig. 8 together with
low temperature data for the  MKA2 mixture (A particles) and the KA mixture.
It is possible to scale the  diffusion constants for the three mixtures as is demonstrated
in Fig. 9, where log(D) for the MKA2 mixture  and the KA  mixture 
are plotted as a function of 1/(T*0.806)(MKA2) and  1/(T*0.714)(KA). A similar  "fragility invariance"
for a system with different repulsive power  potentials ($r^{-n}$) was reported in Ref. \cite{con}.

\section{Discussion}

The highly viscous fluid state below the equilibrium melting point  is found in
many systems ranging from simple organic single-component systems and alloys
to complex biochemical substances and ionic liquids.
Along with the increasing interest in this state of matter
it is important to have models that are not prone to  crystallisation.
Based on the CNT in Sec. II and the simulations of the modified systems in Sec. III we 
estimate that the crystallization time for the standard Kob-Andersen mixture
at $T=0.40$ is of the order 5$\times 10^{7}$ time units ($\approx .1 ms$ Argon units).

The general thermodynamic theory for (equilibrium)
melting and classical nucleation theory
show that an exothermic mixing suppresses crystallization.  On the other hand,  it is the
exponential decreasing self diffusion near the glass transition in highly viscous liquid which
is the dominating factor for  the stabilization of highly viscous liquid (Table I). Theory also
gives an indication
of how to create models that suppress crystallization by cooling.
We have given one simple example by starting from the modified
the Kob-Andersen model, but it is straightforward to
extend the model to include other interactions and  more refined energy functions without loosing  stability
against crystallization. A general recipe of energy functions for stable highly viscous liquids is, however,
difficult to set up, since many mixtures exhibit "reacting systems" with crystals with mixed compositions
where an increased binding-or mixing energy can enhance crystallization of, e.g., A,B crystals.
Both the Kob-Andersen and the Wahnstr{\"o}m models are  examples of this phenomenon.

The new model enhances the relative attraction between unlike species by removing the attraction for
the A,A- and B,B-interactions. Supercooled liquid (WCA-) mixtures with solely repulsive
LJ-potentials have been proposed  before \cite{Evans}, \cite{Chandler},
but we find that both systems studied in Refs. \cite{Evans} and \cite{Chandler}
phase separate and crystallize by cooling. By maintaining the attraction between unlike species,
however, the highly viscous state is stabilized. The new  system 
is faster to simulate than the original KA BLJ mixture.
The new model has the same energy scaling parameter, $\epsilon_{\alpha \beta}$, but a much stronger
violation of the energy rule in the interaction interval $r_{\alpha \beta} \in [2^{1/6} \sigma_{i,i},
\infty]$ where the WCA-potentials are zero. The result of this "cut and shift" is that the B-particles
are  almost "covalently" bounded to the A-particles. Thus   violation of the LB-rule 
gives a recipe for creating fragile and stable supercooled liquid mixtures.

{\it Note added in proof.} Using high-end graphics cards http://www.nvidia.com/cuda we recently performed seven independent
simulations of the KA system with T=0.40. During the runs of 7.4 109 time steps 3.7 107 time units, two out of the seven samples crystallized. The estimated mean
nucleation time  is 1.1 108. This nucleation data are consistent with the extrapolation presented in Table I.

\acknowledgments

The centre for viscous liquid dynamics ``Glass and Time'' is sponsored by the Danish National Research Foundation's (DNRF).

\newpage

 TABLE I: KA and MKA systems  at $T=0.40$ and  $\rho=1.2$.
  \\
  \begin{tabbing}
  \hspace{2.cm}\=\hspace{1.5cm}\=\hspace{3.0cm}\=\hspace{2.cm}\=\hspace{2.5cm}\=\hspace{2.cm}\\
 $System$\> $n$\>  $D_{\textrm{A}}$\>$u_{\textrm{pot,A}}$ \> $\tau$\> $ln(\tau D$)\\
------------------------------------------------------------------------------------------------\\
MKA3\> 3 \> 4.54 $\times 10^{-5}$\> -5.582 \> 2.70 $\times 10^{5}$\> 2.51 \\
MKA2.5\> 2.5 \> $3.64 \times 10^{-5}$\> -5.696 \> 4.92  $\times 10^{5}$\> 2.89 \\
MKA2\> 2\> $1.79 \times 10^{-5}$\> -5.850 \> 1.82  $\times 10^{6}$ \> 3.48  \\
MKA1.5\> 1.5\> $9.99 \times 10^{-6}$\> -6.043 \> 2.97 $\times 10^{6}$ \> 3.46  \\
KA\> 0 \>  9.18 $\times 10^{-7}$\> -6.584 \> $\approx$ 5. $\times 10^{7}$ \> (3.83)\\
 \> \> \> \> (extrapolated) \> \\
\end{tabbing}
------------------------------------------------------------------------------------------------\\
Length unit: $\sigma_{\textrm{AA}}$, time unit:
$\sigma_{\textrm{AA}} \sqrt{m/\epsilon_{\textrm{AA}}} (\approx 2. \times 10^{-12} s$ in Argon units).
\newpage
{\bf  Figure Captions}\\

Figure 1.\\
 Melting point temperature $T_{{\textrm{fus,A}}}$ of the solvent of A particles
 at the external pressure $p$=10 as a function of the particle fraction
 $x_{{\textrm{B}}}$ of solute particles, and in the concentration interval
  $x_{{\textrm{B}}} \in [0, 0.2]$ where the system crystallizes into a crystal of pure (fcc) A-particles.
 (1): With full line is  $T_{{\textrm{fus,A}}}(x_{{\textrm{B}}})$ for an
 ideal mixture and  the Wahnstr{\"o}m-model \cite{wahn} (no visible difference),
 (2): with small dashes is for the MKA2 system ($n=2$) (see Sec. III) ,
 and (3): with long dashes is the present new mixture without like-particle attractions (see Sec. IV).\\

Figure 2.\\
 Radial distribution functions for the solvent-solute particles, $g_{\textrm{AB}}(r)$ and the solvent-solvent
 particles $g_{\textrm{AA}}(r)$ at $T=0.40$ and $\rho$=1.2 and for the four systems with data given in
 Table I. (1) is $g_{\textrm{AB}}(r)$ and (2) is  $g_{\textrm{AA}}(r)$.
 \\

Figure 3.\\

(a) Particle positions projected onto a plane for the MKA2  liquid (the large (A) particles are green, the small (B) particles are black). The system is shown after 1.5 $\mu$s of simulation (Argon units) at T=0.40 and density 1.2 (dimensionless units). After this simulation time the A particles phase separated and formed a large crystal, as is clear in the second figure showing the same configuration with all A particles removed.\\

Figure 4.\\

Schematic illustration of the change in melting point temperature, $T_m(x_{\textrm{B}})$ as a function of the particle fraction, $x_{\textrm{B}}$ by removing the attraction between the A,A-  and B,B pair interactions.
\\
\\

Figure 5.\\

Particle positions of the frozen  system (Sec. IV) after it first was grafted with a small CsCl (A,B) crystal and  at $T=0.30$. The A-particles are green and the B-particles are red. 
(a) The hole system of 1000 particles at the density $\rho=1.2$ and $\xb=0.20$. (b) A central part of the frozen system where the two coexisting crystal forms, fcc (A) and CsCl (AB) clearly can be seen. (The frozen system contains several defects).\\

Figure 6.\\
 Radial distribution functions (1): $g_{\textrm{AB}}(r)$; (2): $g_{\textrm{AA}}(r)$, and 
(3): $g_{\textrm{BB}}(r)$, for the  distribution of particles at
 the density $\rho$=1.2 and the temperature T=0.50 for a mixtures with
 particle fraction $x_{\textrm{A}}$=0.8 of B-particles. Full line is for the MKA2 mixture,
 the dashed curve is for the new mixture.\\

Figure 7.\\
 $\log-\log$ plot of the mean-square displacement for the A-particles of the new BLJ as a function of time
 (in unit $\sigma_{\textrm{AA}} \sqrt{m/\epsilon_{\textrm{AA}}} \approx 2. \times 10^{-12} s)$ and for different temperatures (from the left): T=1.00, 0.40, 0.35, 0.325, 0.30, 0.275 and 0.25. respectively.\\

Figure 8.\\

An Arrhenius plot, $\log$ D(1/T), of the self diffusion constant D. (1): With filled squares
and connected with dash-dots is for the
KA mixture. (2): 
With + is D(A) for the MKS2 mixture, and (3): the points given by $\times$ and connected with dashes is D(A) for the new binary mixture (see III B). (4): The diffusion constants, D(B) for the smaller B-particles in the  new binary mixture are shown with $\star$.\\

Figure 9.\\
 With full line and $\times$ is the $\log$D(1/T) for the new mixture  from Fig.8. With dashes are the corresponding scaled MKA2 data $\log$(D)$(\frac{1}{0.806T})$, and with dot-dashes are the corresponding  KAI data $\log$(D)$(\frac{1}{0.714T})$.

\newpage
\begin{figure}\begin{center}
\includegraphics[height=8cm]{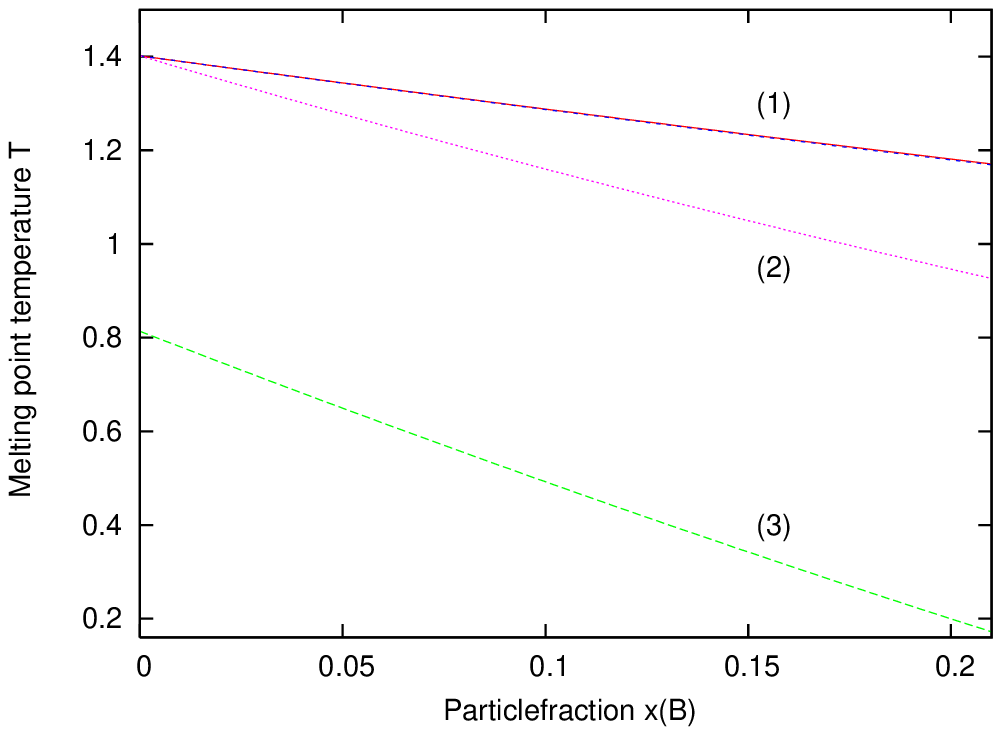}
\caption{}
\end{center}\end{figure}
\begin{figure}\begin{center}
\includegraphics[height=7cm]{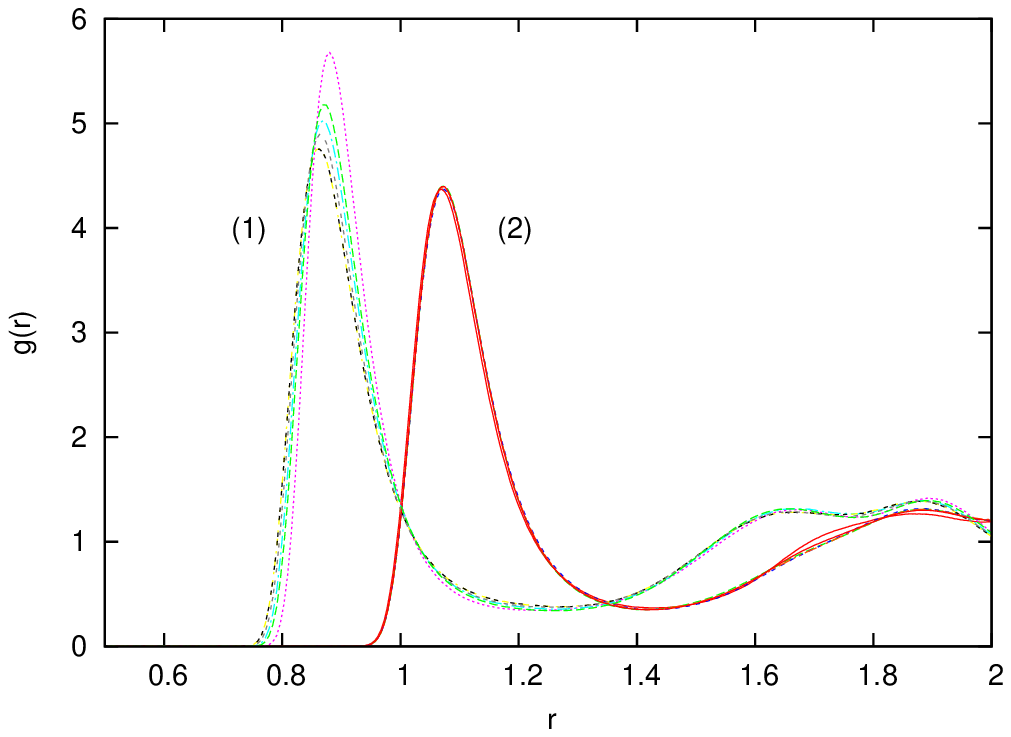}
\caption{}
\end{center}\end{figure}
\begin{figure}\begin{center}
\includegraphics[height=8cm]{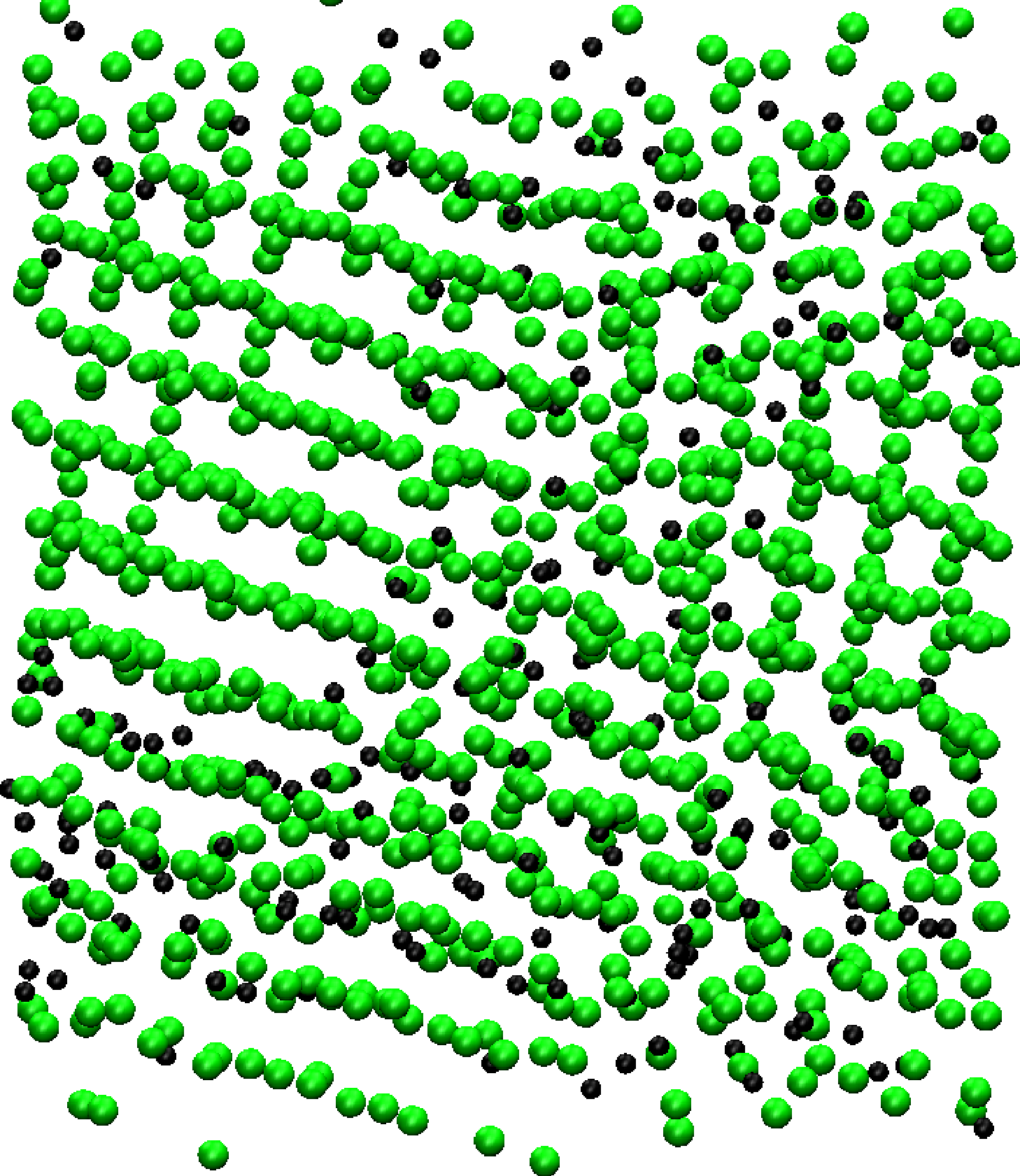}
\includegraphics[height=8cm]{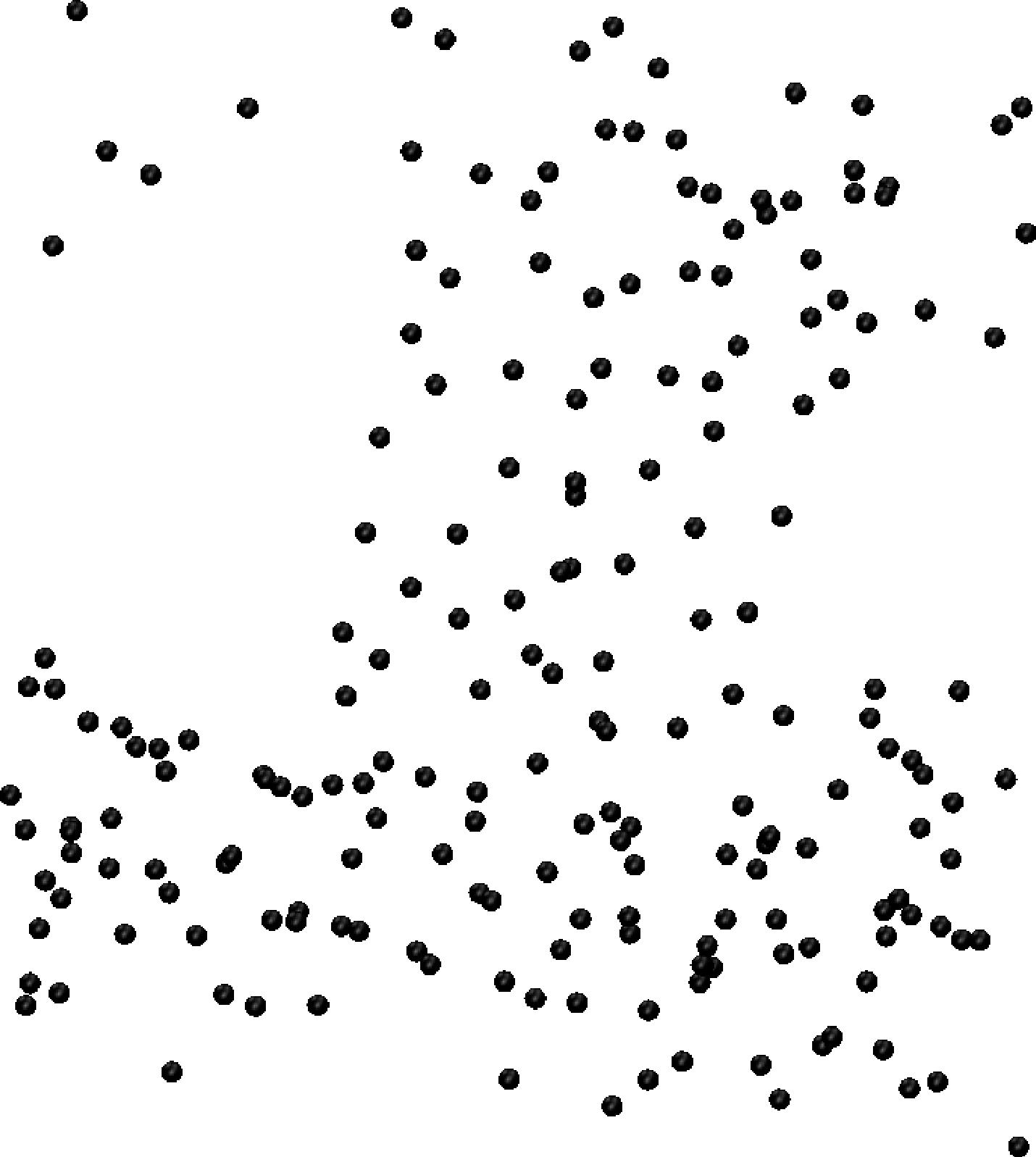}
\caption{}
\end{center}\end{figure}
\begin{figure}\begin{center}
\includegraphics[height=8cm]{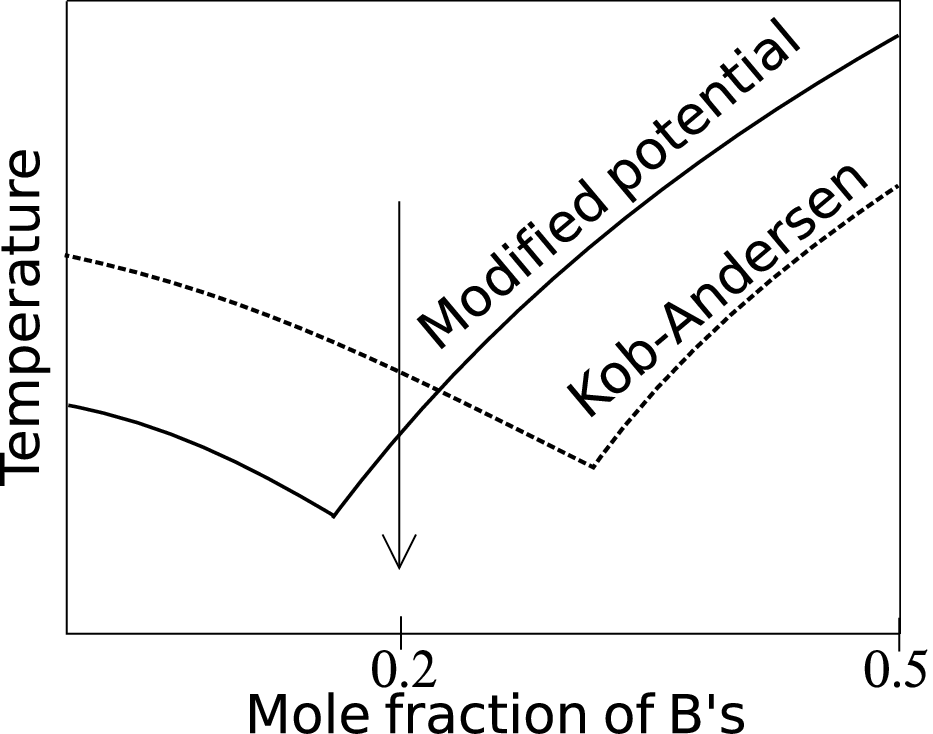}
\caption{}
\end{center}\end{figure}

\begin{figure}\begin{center}
\includegraphics[height=8cm]{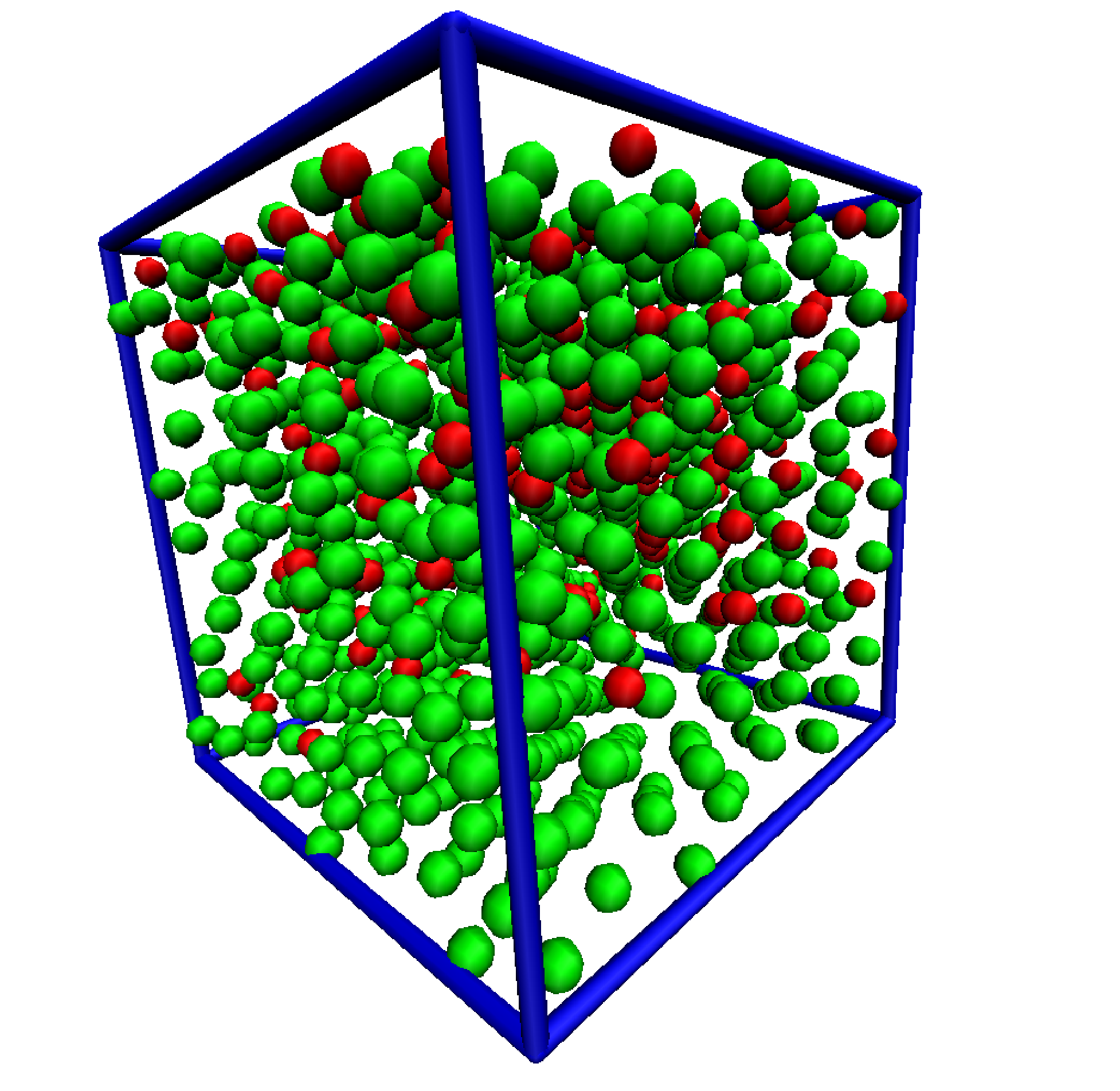}
\includegraphics[height=8cm]{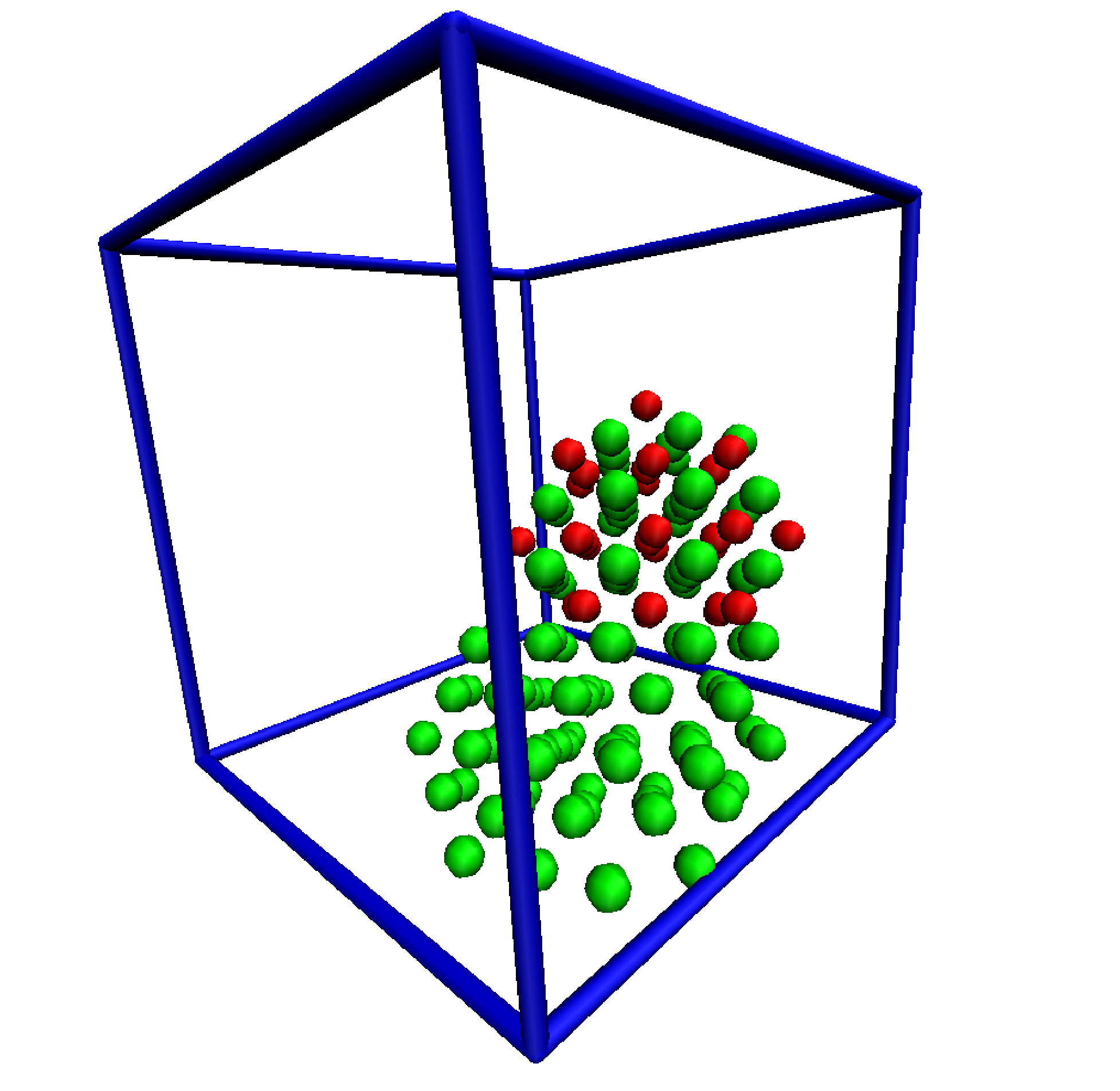}
\caption{}
\end{center}\end{figure}
\begin{figure}\begin{center}
\includegraphics[height=8cm]{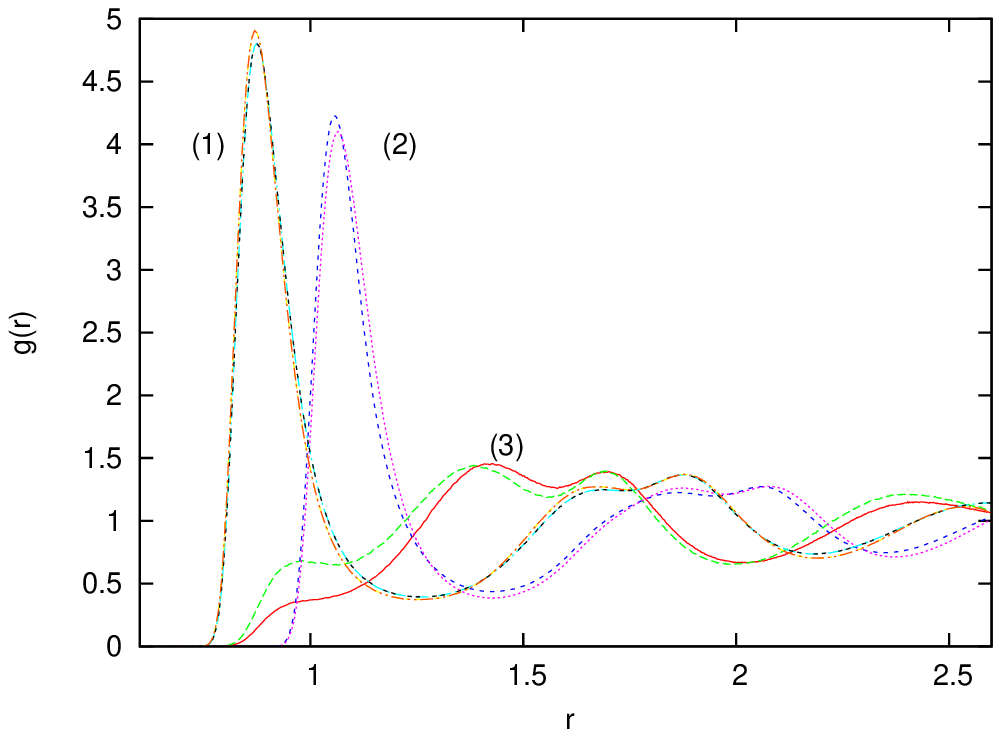}
\caption{}
\end{center}\end{figure}
\begin{figure}\begin{center}
\includegraphics[height=8cm]{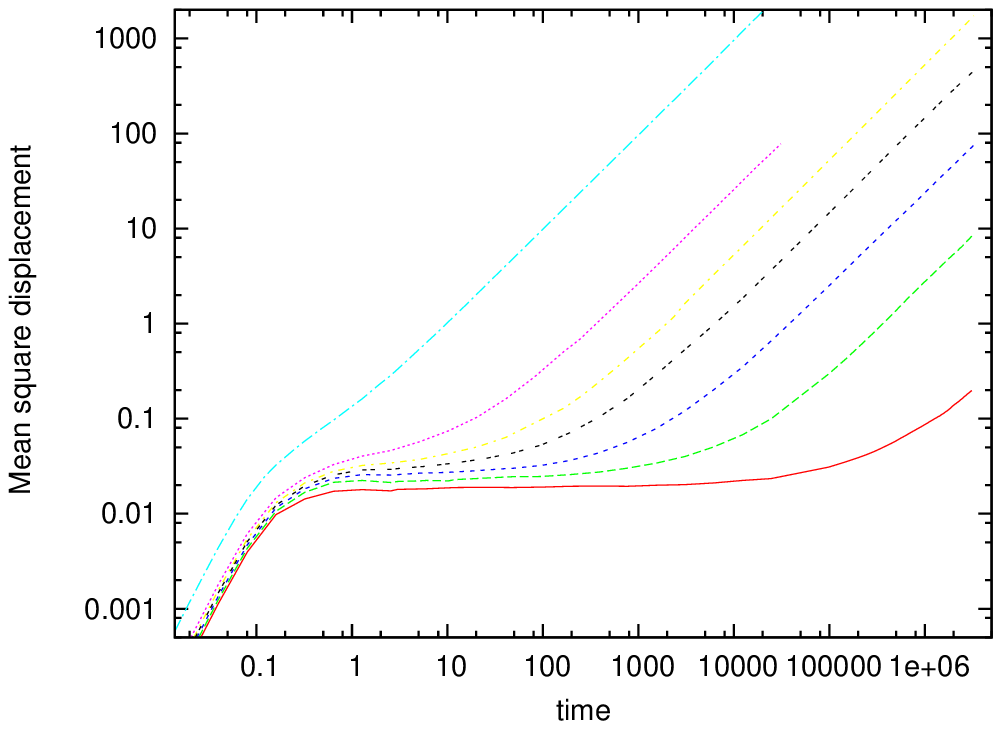}
\caption{}
\end{center}\end{figure}
\begin{figure}\begin{center}
\includegraphics[height=8cm]{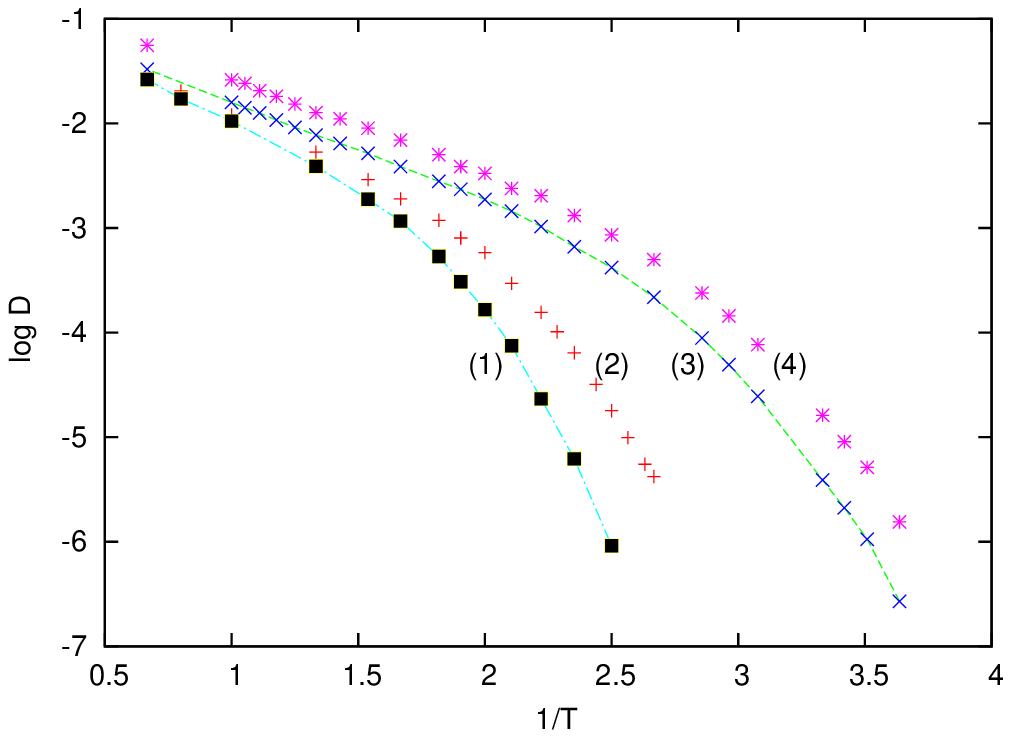}
\caption{}
\end{center}\end{figure}
\begin{figure}\begin{center}
\includegraphics[height=9cm]{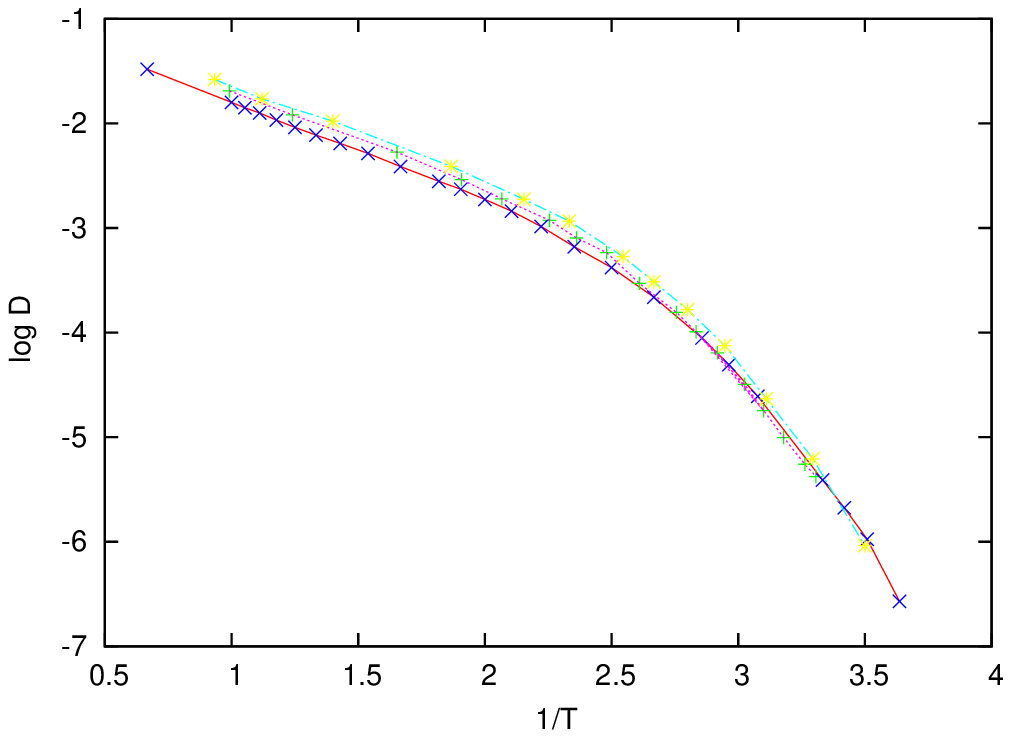}
\caption{}
\end{center}\end{figure}

\end{document}